# Determination of the Number of Graphene Layers: Discrete Distribution of the Secondary Electron Intensity Derived from Individual Graphene Layers

Hidefumi Hiura[1, 2 *], Hisao Miyazaki[2, 3], and Kazuhito Tsukagoshi[2, 3]

[1]*Green Innovation Research Laboratories, NEC Corporation, Tsukuba, Ibaraki 305-8501, Japan,*

[2]*International Center for Materials Nanoarchitectonics (MANA), National Institute for Materials Science (NIMS), Tsukuba, Ibaraki 305-0044, Japan*

[3]*CREST, Japan Science and Technology Agency, Kawaguchi, Saitama 305-0044, Japan*

## ABSTRACT

Using a scanning electron microscope, we observed a reproducible, discrete distribution of secondary electron intensity stemming from an atomically thick graphene film on a thick insulating substrate. The discrete distribution made it possible to uniquely relate the secondary electron intensity to the number of graphene layers. Furthermore, we found a distinct linear relationship between the relative secondary electron intensity from graphene and the number of layers, provided a low primary electron acceleration voltage was used. Based on these observations, we propose a practical method to determine the number of graphene layers in a sample. This method is superior to the conventional optical method in its capability to characterize graphene samples with sub-micrometer squares in area on various insulating substrates.

*E-mail address: h-hiura@bq.jp.nec.com (or HIURA.Hidefumi@nims.go.jp)

Graphene has attracted broad attention since the discovery of an electrical field effect in atomically thick graphene[1]. The various unique properties of graphene have created high expectations for a number of applications in emerging technologies[2–5]. In particular, charge carriers in graphene behave uniquely; individual transport phenomena have been found in monolayer ($1L$)[1,6–8], bilayer ($2L$)[9–11], trilayer ($3L$)[12], and multilayer graphene[13].

Prior to these discoveries in graphene, electrical transport in very thin graphite crystals was reported by Mizushima *et al.* in 1971[14]. Since their graphite crystals were over 10 nm thick, it is difficult to classify them as graphene, and at that time there was no need to count the layers in the samples. Nevertheless, graphene layers are atomically thick, and the overall electronic properties of a sample are determined by the number of layers present. Therefore, it is of great importance to be able to experimentally determine the number of graphene layers in a sample, particularly for device fabrication. To date, several methods of measuring the number of graphene layers have been reported, such as optical microscopy[15–18], atomic force microscopy[19,20], Raman microscopy[21–23], low-energy electron microscopy[24], and Auger electron spectroscopy[25]. Among these, optical and Raman microscopy are commonly used because they are the most convenient and reliable. However, neither of these methods is applicable to sub-micrometer areas, due to the diffraction limits of light. In addition, the former is practical only for graphene on a silicon substrate covered by $SiO_2$ with a specific thickness, namely, ~100 or ~300 nm[15–17]. The latter method has the major drawback that the moderate laser irradiation could cause structural degradation of the graphene[23]. Scanning electron microscopy (SEM) is also frequently used, but is subsidiary to the above-mentioned layer counting techniques for graphene, because no decisive SEM technique has yet been established to identify the number of graphene layers. Here, we report that individual graphene layers, from one to several, can be distinguished on the basis of the discretely distributed secondary electron intensity in a conventional SEM. We found that the secondary electron intensity of graphene relative to that of the substrate decreases proportionally as the number of layers increases, enabling the measurement of the number of graphene layers on various insulating substrate.

Graphene samples were prepared by the so-called glue tape method[6, 7]. Graphene or thin graphite flakes were mechanically exfoliated from natural graphite with a pair of glue tapes, and were then attached to the surface of one of three different substrates: silicon with a 300-nm thermal silicon dioxide surface layer ($SiO_2$/Si), sapphire, or mica. Using a normal optical microscope, the number of graphene layers on the $SiO_2$/Si substrates were predetermined by reflectance measurements[15–17], while those on the sapphire or mica substrates were determined by transmittance measurements[18]. SEM images of graphene were observed in vacuo using a standard SEM system (VE-9800, KEYENCE) equipped with an outer detector to collect the secondary electrons accelerated to an energy of 10 keV.

Figures 1 (a)–(h) show SEM images of a graphene flake on the surface of an $SiO_2$/Si substrate, observed with various primary electron acceleration voltages $V_{acc}$ from 0.5 to 20 kV. As indicated in the optical microscope image, Fig. 1 (i), this graphene flake was composed of one to several layers of graphene ($L$ = 1, 3, 4, 5, 6, 7, and 8) at the periphery, and a graphite region ($L \gg 10$) in the center. There are three important points regarding Fig. 1. First, depending on $V_{acc}$, the contrast of graphene to the $SiO_2$ surface changed drastically. The dependence on $V_{acc}$ was more prominent in the SEM images with fewer layers of graphene; the fewer the number of graphene layers, the larger the change in SEM contrast. Second, each individual graphene layer can be clearly distinguished when observed with a $V_{acc}$ of 0.5 to 2.0 kV. In particular, a few layers of graphene gave a brighter image than the $SiO_2$ surface when observed in this range, thereby making it possible to identify them more easily. The bright SEM contrast of a few graphene layers is reminiscent of single-walled carbon nanotubes observed at a low $V_{acc}$ of around 1 kV[26–28]. For example, the contrast of nanotubes on an $SiO_2$ surface is highest at $V_{acc}$ = 1–2 kV; dropping significantly for $V_{acc}$ > 3 kV[26]. This is in accord with our observations of graphene. Considering the significantly different morphology and dimensionality between graphene and nanotubes, it is rather surprising that they behave in a similar manner when observed by SEM. Third, as $V_{acc}$ was increased above 3.0 kV, all of the graphene layers become darker than the $SiO_2$ surface. Finally, when observed at $V_{acc}$ = 20.0 kV, there was no apparent difference in SEM contrast between the individual graphene layers, as seen in Fig. 1 (h).

Next, we extracted numerical values representing the contrast from some of the SEM images of graphene, and plotted them as a function of $V_{acc}$ for graphene with $L$ = 1, 2, 4, 6, and 8, and for graphite with L » 10, as shown in Fig. 2. Hereafter, contrast was defined as the relative intensity in the presence or absence of graphene (C = $(I – I_0)/I_0$, $C$: contrast, $I$: the intensity of the secondary electron, reflected light, or transmitted light with graphene, $I_0$: the intensity of the secondary electron, reflected light, or transmitted light without graphene). On one hand, the SEM contrast curves changed gradually from single-layer graphene ($L$ = 1) to graphite ($L$ » 10). On the other hand, the curves from 1$L$ to 8$L$ of graphene shared a common profile: a peak around 1.0 kV with a plateau between 3.0 and 20.0 kV. Similar trends were observed for graphene, on both sapphire and mica substrates. According to the literature[29], the secondary electron yield for $SiO_2$ decreases monotonically, as long as $V_{acc}$ > 0.5 kV. Combining this information with our observations, we believe that the curve profile shown in Fig. 2 is intrinsic to graphene and is related to the secondary electron yield of graphene. The more important aspect of Fig. 2 is that every curve lies at a regular interval, in other words, the SEM contrast linearly decreases with an increasing number of graphene layers at any $V_{acc}$ we used. The only exception is in the case of $L = 1$. This linear relationship can be used to count graphene layers as shown next.

The main panel of Fig. 3 depicts the relationships between the SEM contrast of graphene on an $SiO_2$/Si substrate and the number of graphene layers, measured with varying $V_{acc}$. The SEM contrast was determined in the same manner as mentioned in the previous section, while the number of layers was determined by optical measurements. Even when $V_{acc}$ was varied from 0.5 kV to 10 kV, each set of data points formed a straight line with a negative slope, except for $L = 1$. As shown in the inset of Fig. 3, a similar linearity was also observed for graphene on sapphire and mica. The linear relationship can be explained in terms of attenuation of the secondary electrons by the graphene layers; the secondary electrons should be attenuated at a rate proportional to the number of graphene layers they pass through. Another possible explanation for the linearity may be found in the dependence of the work function of graphene on the number of layers, since a lower work function, in general, leads to a higher secondary electron yield[30]. Hibino *et al.* reported that the work function of graphene varies linearly from ~4.3 eV for $L = 1$ to ~4.6 eV for $L = 4$, but is saturated above $L = 4$[31]. Thus, the variable work function alone cannot account for the linear SEM contrast relationship from $1L$ to $8L$. As shown in Fig. 3, the deviation from linearity for $L = 1$ developed independent of both $V_{acc}$ and the substrate type, indicating that it is peculiar to 1$L$ graphene. The origin of this peculiarity is unclear, but could arise because, unlike multilayer graphene, only a 1$L$ graphene both receives and emits the secondary electrons. Regardless of their physical origin, the brighter SEM images of 1$L$ graphene are easy to identify. Figure 3 shows that the slope of the line varied depending on the experimental conditions, but was unique for a given set of conditions, such as $V_{acc}$ and substrate type. Therefore, once a calibration line is determined by observing SEM images of a standard sample, the number of layers in an unknown graphene sample can be determined, provided the same conditions are applied. In putting the SEM counting method to practical use, the choice of $V_{acc}$ is important. A range of $V_{acc}$ should be chosen to obtain the steeper line slope and brighter contrast to make it easier to find and discriminate individual graphene layers. According to these criteria, a range of $V_{acc}$ between 0.5 and 1.5 kV is suitable for the SEM counting method, and $V_{acc}$ = 1.0 kV is optimal. This preferred voltage was independent of which substrate was used.

Figure 4 shows a comparison of optical and SEM methods for determining the number of graphene layers on various substrates, (a) $SiO_2$/Si, (b) mica, and (c) sapphire. The optical images were taken via reflection for (a), and transmission for (b) and (c), while all SEM images were observed at $V_{acc}$ = 1 kV, the preferred value for layer counting, as discussed above. Both optical and SEM methods followed virtually the same procedure, which was as follows. First, an image of some graphene layers was taken and converted to a contrast histogram by image processing. Second, the contrast axis of the histogram was converted to the number of graphene layers, by simply dividing the contrast by the contrast difference per layer. In the optical results shown in (a) to (c), the contrast difference per layer was −5.29%, −2.26%, and −2.29%, respectively, values which were consistent with the literature ones[15−18]. For the SEM results shown in (a) to (c), the contrast difference per layer was −5.05%, −2.65%, and −6.11%, respectively, the value of which were predetermined from the slope of the calibration line, as shown in the inset of Fig. 3. In both methods, the absolute value of the contrast difference per layer was large enough to discriminate individual graphene layers in the image with the naked eye. This is supported by the emergence of discrete peaks in each histogram, as shown in Fig. 4. Finally, the number of graphene layers was determined from the peak positions. As a result, for the cases (a) to (c), 1*L*, 3*L*, and 5*L* graphene; 1*L*, 2*L*, 4*L*, and 8*L* graphene; and 2*L*, 4*L*, and 9*L* graphene, respectively, were observed in the dotted areas by both optical and SEM methods, demonstrating that the SEM method is equally as valid as the optical method. The SEM method is no less advantageous than previously reported methods. For example, it is non-destructive, since it requires only a low $V_{acc}$, of ∼1 kV. Because SEM has a much higher spatial resolution than that of optical microscopy, it can be utilized for very small samples less than 1 μm$^2$ in area, making it a useful technique to assist in the fabrication of graphene nanodevices. Furthermore, the method can be applied to graphene on almost any insulating substrate. Because of these advantages, we anticipate that this method will serve as a versatile tool for determining the number of graphene layers in a variety of situations.

In conclusion, we observed graphene on several insulating substrates by SEM, and found a reproducible discrete distribution of secondary electron intensity derived from individual graphene layers, making it possible to discriminate individual graphene layers when using low primary electron acceleration voltages. Furthermore, we found a linear relationship between the SEM contrast of graphene and the number of graphene layers, and demonstrated that this linearity enables SEM to be used to easily count graphene layers in a sample. In particular, our SEM counting method will be useful to characterize nanometer-sized graphene samples, which are too small to be observed using the current optical methods.

Acknowledgement

This work was partly supported by the Japan Science and Technology Agency.

Figure 1. SEM images of graphene ($L$ = 1, 3, 4, 5, 6, 7, and 8) and graphite flakes ($L$ » 10) on the surface of a $SiO_2$/Si substrate, measured at several different primary electron acceleration voltages $V_{acc}$. Each SEM image was observed under the same conditions, except that $V_{acc}$ was set to 0.5 kV (a), 0.8 kV (b), 1.0 kV (c), 1.4 kV (d), 2.0 kV (e), 3.0 kV (f), 5.0 kV (g), or 20.0 kV (h). For comparison, a reflected optical image of the same graphene flakes is depicted in (i). The number of graphene layers is indicated in the figures.

Figure 2. Dependence of SEM contrast on the primary electron acceleration voltage $V_{acc}$ for 1$L$, 2$L$, 4$L$, 6$L$, and 8$L$ graphene, and » 10$L$ graphite on the surface of a $SiO_2$/Si substrate. The contrast was defined as the relative intensity in the presence vs. absence of graphene.

Figure 3. Relationships between the SEM contrast of graphene to the substrate surface and the number of graphene layers. The contrast was obtained from SEM images, while the number of layers was determined by optical measurements. The relationship in the main panel was measured for graphene with various layer counts on a $SiO_2$/Si substrate. $V_{acc}$ was maintained at 0.5 kV, 1.0 kV, 1.5 kV, 3.0 kV, or 10 kV. The inset shows a comparison of the SEM contrast versus the number of graphene layers, obtained at $V_{acc}$ = 1.0 kV using $SiO_2$/Si (a), mica (b), and sapphire (c) substrates.

Figure 4. Comparison of the counting of layers by optical microscopy and by SEM for graphene on various substrates: $SiO_2$/Si (a), mica (b), and sapphire (c). For each figure from (a) to (c), the upper and lower figures show optical and SEM images, respectively, of the same graphene sample, along with a histogram of the distribution of graphene layers within the rectangular area indicated by a dotted line. The optical images were taken through reflected light for (a), and transmitted light for (b) and (c). All SEM images were observed at $V_{acc}$ = 1 kV. For each histogram, the vertical axis on the left is marked by a linear scale showing the contrast of graphene to the substrate, while the axis on the right is marked by a linear scale representing the number of graphene layers. An asterisk at zero contrast indicates a substrate peak. The scale bars in the upper images are 5 μm long.

# Figure 1

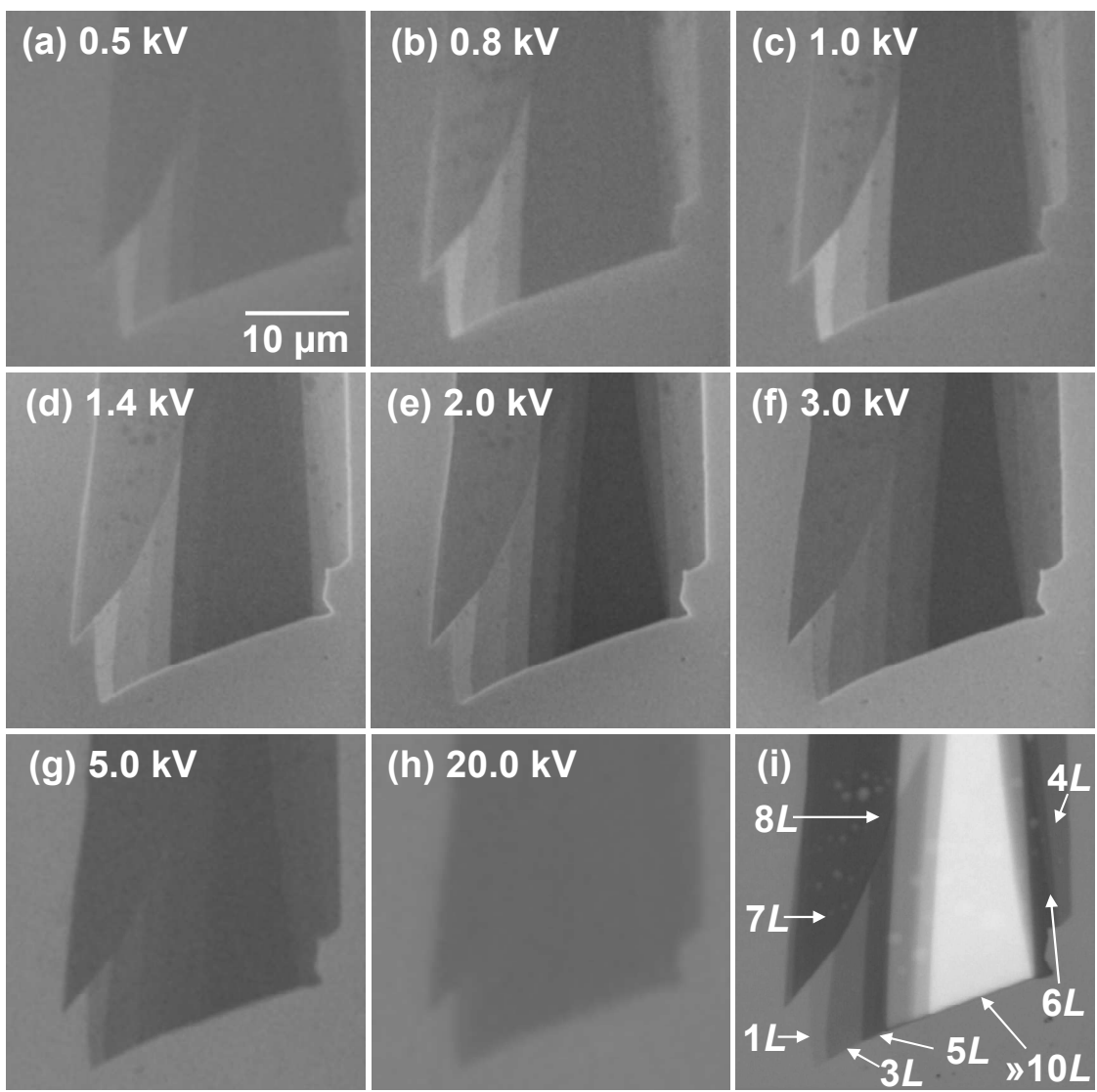

**Figure 2**

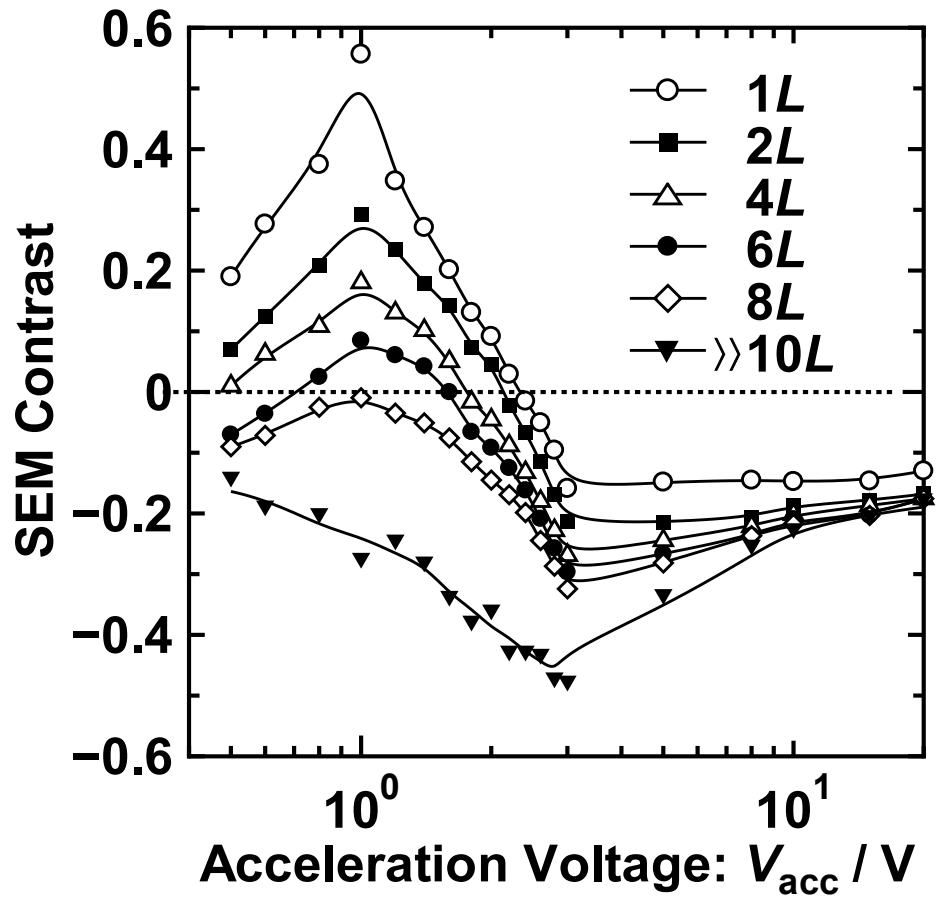

# Figure 3

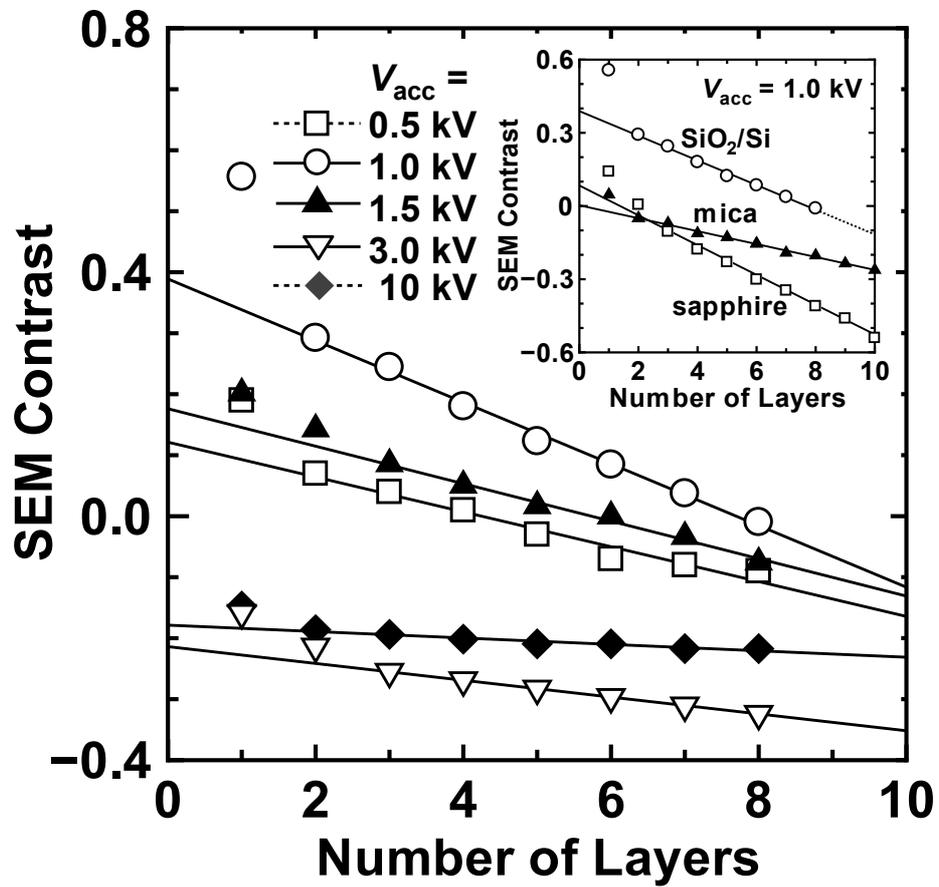

# Figure 4

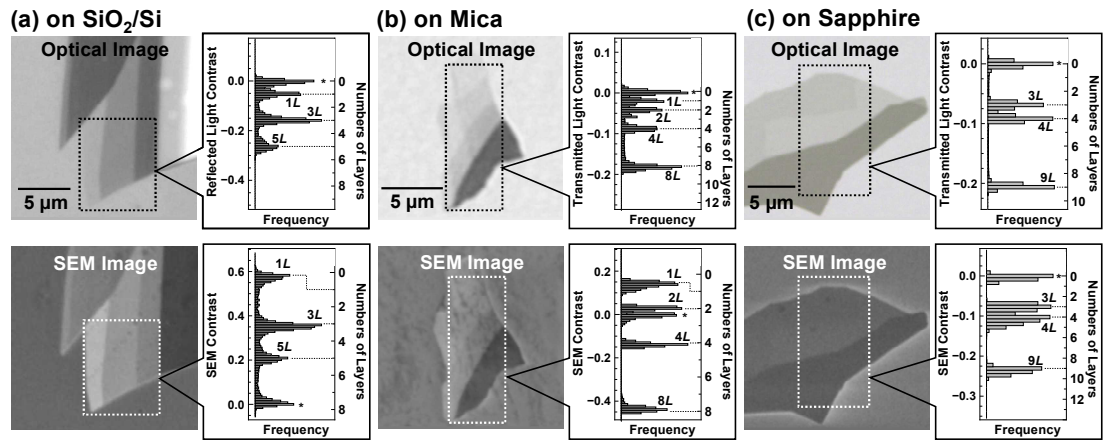